\documentclass[a4paper,10pt,twoside]{cpc-hepnp}

\usepackage{multicol}
\usepackage{graphicx}
\usepackage{booktabs}
\usepackage{amssymb,bm,mathrsfs,bbm,amscd}
\usepackage[tbtags]{amsmath}
\usepackage{lastpage}

\begin{document}

\fancyhead[co]{\footnotesize B. S. Zou : Complementary studies on
$N^*$ from $e^+e^-$, $pp$ and $p\bar p$ collisions}

\footnotetext[0]{Received 28 July 2009}

\title{Complementary studies on $N^*$ from $e^+e^-$, $pp$ and $p\bar p$ collisions\thanks{Supported
by National Natural Science Foundation of China (10875133,10821063),
Chinese Academy of Sciences (KJCX3-SYW-N2) and Ministry of Science
and Technology of China (2009CB825200) }}

\author{%
      B. S. Zou$^{1,2;1)}$\email{zoubs@ihep.ac.cn}%
}
\maketitle

\address{%
1~(Institute of High Energy Physics, Chinese Academy of Sciences, Beijing 100049, China)\\
2~(Theoretical Physics Center for Science Facilities, Chinese Academy of Sciences, Beijing 100049, China)\\
}

\begin{abstract}
Complementary to the conventional experimental studies on $N^*$ from
$\pi N$ and $\gamma^{(*)}N$ reactions, the $e^+e^-$, $pp$ and $p\bar
p$ collisions can give novel insights into these $N^*$ resonances.
While the $e^+e^-$ collisions through production and decay of vector
charmonium $\psi$ provide a nice isospin filter for a simultaneously
study of $N^*$, $\Delta^*$, $\Lambda^*$, $\Sigma^*$ and $\Xi^*$, the
$pp$ collisions should be the best place for producing those
$\Delta^{*++}$ with large coupling to $\rho^+p$ though $pp\to
n\Delta^{*++}$ reaction, and the $p\bar p$ collisions should be the
best place for looking for those $N^*$ with large coupling to
$\sigma N$.
\end{abstract}



\begin{multicols}{2}

\section{Introduction}

There are two well-known problems for the classical $3q$ constituent
quark models. The first one is the mass reverse problem for the
lowest excited states. In the simple 3q constituent quark model, the
lowest spatial excited baryon is expected to be a ($uud$) $N^*$
state with one quark in orbital angular momentum $L=1$ state, and
hence should have negative parity. Experimentally~\cite{PDG08}, the
lowest negative parity $N^*$ resonance is found to be $N^*(1535)$,
which is heavier than two other spatial excited baryons :
$\Lambda^*(1405)$ and $N^*(1440)$. In the classical 3q constituent
quark model, the $\Lambda^*(1405)$ with spin-parity $1/2^-$ is
supposed to be a ($uds$) baryon with one quark in orbital angular
momentum $L=1$ state and about 130 MeV heavier than its $N^*$
partner $N^*(1535)$; the $N^*(1440)$ with spin-parity $1/2^+$ is
supposed to be a ($uud$) state with one quark in radial $n=1$
excited state and should be heavier than the $L=1$ excited ($uud$)
state $N^*(1535)$, noting the fact that for a simple harmonic
oscillator potential the state energy is $(2n+L+3/2)\hbar\omega$. So
for these three lowest spatial excited baryons, the classical quark
model picture is already failed. The second problem is that in many
of its forms it predicts a substantial number of `missing $N^*$
states' around 2 GeV/$c^2$, which have not so far been
observed~\cite{Capstick1}. Since the more number of effective
degrees of freedom the more predicted number of excited states, the
`missing $N^*$ states' problem is argued in favor of the diquark
picture which has less degree of freedom and predicts less $N^*$
states~\cite{diquark}. For example, in diquark models, the two
quarks forming the diquark are constrained to be in the relative
S-wave, and hence cannot combine the third quark to form
(20,$1^+_2$)-multiplet baryons. Experimentally, not a single
(20,$1^+_2$)-multiplet baryon has been identified yet~\cite{PDG08}.
However, non-observation of these `missing $N^*$ states' does not
necessarily mean that they do not exist. In the limit that the
$\gamma$ or $\pi$ couples to one quark in the nucleon in the $\gamma
N$ or $\pi N$ reactions, the (20,$1^+_2$)-multiplet baryon cannot be
produced~\cite{Zhaoq}. Considering higher order effects, they may
have weak coupling to $\pi N$ and $\gamma N$, but may be too weak to
be observed by presently available $\pi N$ and $\gamma N$
experiments~\cite{Capstick1,Zhaoq}.

To solve the mass order reverse problem, it seems necessary to go
beyond the simple quenched $3q$ quark models. In fact the spatial
excitation energy of a quark in a baryon is already comparable to
pull a $q\bar q$ pair from the gluon field. Even for the proton, the
well established $\bar{d}/\bar{u}$ asymmetry with the number of
$\bar d$ more than $\bar u$ by an amount $\bar d-\bar u\approx
0.12$~\cite{Garvey} demands its 5-quark components to be at least
$12\%$. The 5-quark components can be either in the form of meson
cloud, such as $n(udd)\pi^+(u\bar d)$, or in other forms of quark
correlation, such as penta-quark configuration $[ud][ud]\bar d$ with
$[ud]$-diquark correlation. In either meson cloud model or
penta-quark model, the mass order reverse problem of $N^*(1535)$ and
$\Lambda^*(1405)$ can be easily explained. In the meson cloud
models~\cite{kaiser,jido}, the $N^*(1535)$ is explained as a
$K\Lambda$-$K\Sigma$ quasi-bound state while $\Lambda^*(1405)$ is a
dynamically generated state of coupled $KN$-$\Sigma\pi$ channels. In
the penta-quark models~\cite{Helminen,zhu,Zou-Nstar07}, the
$N^*(1535)$ is mainly a $[ud][us]\bar s$ state while
$\Lambda^*(1405)$ is mainly a $[ud][sq]\bar q$ state with $q\bar
q=(u\bar u+d\bar d)/\sqrt{2}$.

These unquenched models give interesting predictions for the SU(3)
partners of the $\Lambda^*(1405)$ and $N^*(1535)$. For example, the
penta-quark models~\cite{zhu} predict a $\Sigma^*(1/2^-)$ resonance
with a mass around $\Sigma^*(1385)-3/2^+$ and a $\Xi^*(1/2^-)$
around $\Xi^*(1530)-3/2^+$. These predicted states are still
`missing' from PDG list~\cite{PDG08}. However, possible evidence for
their existence in $J/\psi$ decays~\cite{Zou-Charm06} and
$K^-p\to\Lambda\pi^+\pi^-$ reaction~\cite{wujj} has recently been
pointed out.

To look for these `missing' baryon resonances to establish correct
picture for the baryon structure, the scheduled high statistics
$\gamma p$ and $Kp$ experiments are necessary. Here we want to show
that the $e^+e^-$, $pp$ and $p\bar p$ collisions could also play
unique complementary role and should be explored.

\section{$N^*$ from  $e^+e^-\to\psi\to\bar NN^*$}

The $J/\psi$ and $\psi'$ experiments at BES provide an excellent
place for studying excited nucleons and hyperons -- $N^*$,
$\Lambda^*$, $\Sigma^*$ and $\Xi^*$ resonances~\cite{Zou2}.
Comparing with other facilities, the BES baryon program has
advantages in at least three obvious aspects:

(1) For the $c\bar c\to\bar NN\pi$ and $\bar NN\pi\pi$ processes,
the $\pi N$ and $\pi\pi N$ systems are expected to be dominantly
isospin 1/2 due to that the isospin-conserving three-gluon
annihilation of the constituent c-quarks dominates over the isospin
violating decays via intermediate photon for the baryonic final
states, while $\pi N$ and $\pi\pi N$ systems from $\pi N$ and
$\gamma N$ experiments are mixture of isospin 1/2 and 3/2 with
similar strengths, and hence suffer difficulty on the isospin
decomposition;

(2) $\psi$ mesons decay to baryon-antibaryon pairs through three or
more gluons. It is a favorable place for producing hybrid (qqqg)
baryons, and for looking for some ``missing" $N^*$ resonances, such
as members of possible (20,$1^+_2$)-multiplet baryons, which have
weak coupling to both $\pi N$ and $\gamma N$, but stronger coupling
to $g^3N$;

(3) Not only $N^*$, $\Lambda^*$, $\Sigma^*$ baryons, but also
$\Xi^*$ baryons with two strange quarks can be studied. Many
QCD-inspired models~\cite{Capstick1} are expected to be more
reliable for baryons with two strange quarks due to their heavier
quark mass. More than thirty $\Xi^*$ resonances are predicted where
only two such states are well established by experiments. The theory
is totally not challenged due to lack of data.

\begin{center}
\includegraphics[width=7.5cm]{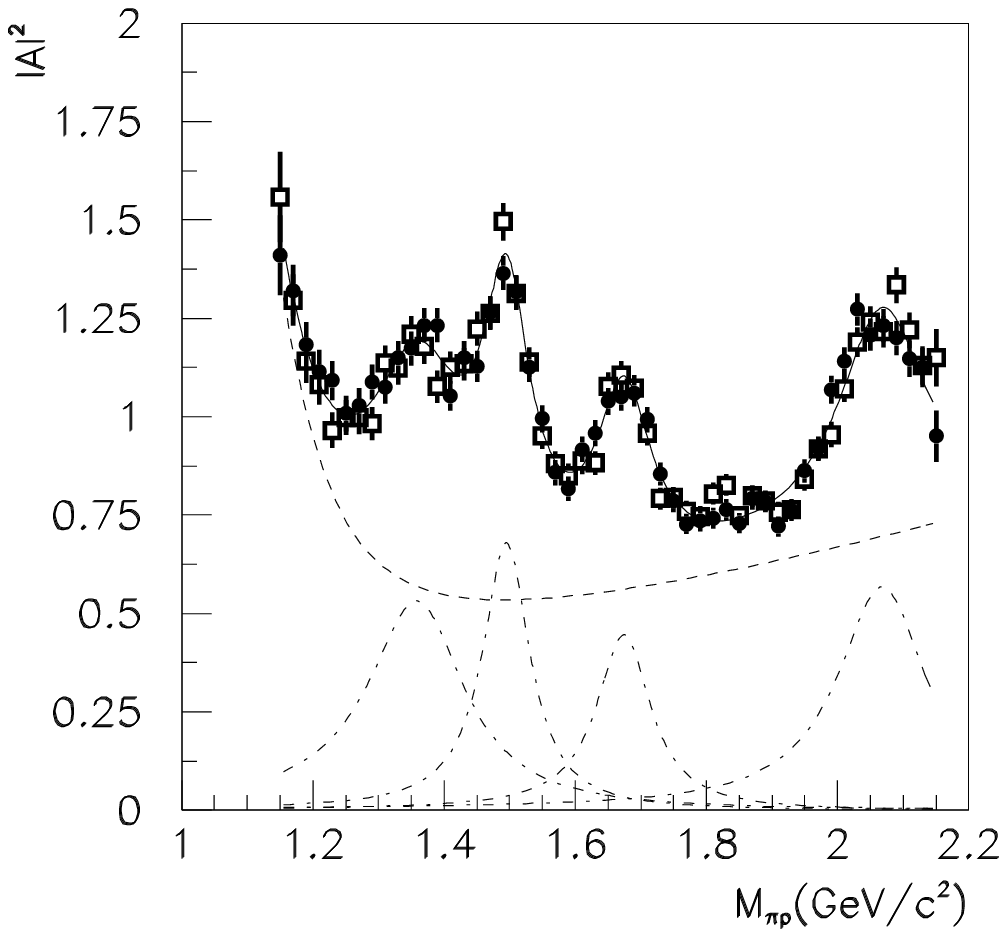}
\figcaption{\label{fig3.2}   Data divided by MC phase space vs
$p\pi$ invariant mass for $J/\psi\to \bar p\pi^-\bar n$ (solid
circle) and $\bar p\pi^+n$ (open square) from Ref.[14]. }
\end{center}

A typical example showing the isospin and spin filter effect is
given by the study of $J/\psi\to p\bar n\pi^- + c.c.$
channel~\cite{bes2}. The data vs $p\pi$ invariant mass divided by
Monte Carlo phase space including the detection efficiency are shown
in Fig.~\ref{fig3.2}. At low $p\pi$ invariant mass, the tail from
nucleon pole term, expected from theoretical
considerations~\cite{Okubo,Liang2}, is clearly seen. There are
clearly four peaks around 1360 MeV, 1500 MeV, 1670 MeV and 2065 MeV.
Note that the well known first resonance peak ($\Delta(1232)$) in
$\pi N$ and $\gamma N$ scattering data does not show up here due to
the isospin filter effect of the $J/\psi$ decays. While the two
peaks around 1500 MeV and 1670 MeV correspond to the well known
second and third resonance peaks observed in $\pi N$ and $\gamma N$
scattering data, the two peaks around 1360 MeV and 2065 MeV have
never been observed in $\pi N$ invariant mass spectra before. The
one around 1360 MeV should be from $N^*(1440)$ which has a pole
around 1360 MeV~\cite{PDG08,Manley,Dytman} and which is usually
buried by the strong $\Delta$ peak in $\pi N$ and $\gamma N$
experiments; the other one around 2065 MeV may be due to the long
sought ``missing" $N^*$ resonance(s). For the decay $J/\psi\to\bar
NN^*(2065)$, the orbital angular momentum of $L=0$ is much preferred
due to the suppression of the centrifugal barrier factor for $L\geq
1$. For $L=0$, the spin-parity of $N^*(2065)$ is limited to be
$1/2+$ and $3/2+$. This may be the reason that the $N^*(2065)$ shows
up as a peak in $J/\psi$ decays while only much broader structures
show up for $\pi N$ invariant mass spectra above 2 GeV in $\pi N$
and $\gamma N$ production processes~\cite{Gaohy} which allow all
$1/2\pm$, $3/2\pm$, $5/2\pm$ and $7/2\pm$ $N^*$ resonances around
2.05 GeV to overlap and interfere with each other there. A simple
Breit-Wigner fit~\cite{bes2} gives the mass and width for the
$N^*(1440)$ peak as $1358\pm 6 \pm 16$ MeV and $179\pm 26\pm 50$
MeV, consistent perfectly with the PDG pole value for the
$N^*(1440)$, {\sl i.e.}, $1365\pm 15$ MeV and $190\pm 30$ MeV,
respectively.  For the new $N^*$ peak above 2 GeV the fitted mass
and width are $2068\pm 3^{+15}_{-40}$ MeV and $165\pm 14\pm 40$ MeV,
respectively. A partial wave analysis indicates that the $N^*(2065)$
peak contains both spin-parity $1/2+$ and $3/2+$
components~\cite{bes2}. Very recently, a detailed partial wave
analysis of the $J/\psi \to p\bar p \pi^0$ channel concludes besides
a $1/2^+$ $N^*(2100)$ a ${3/2}^+$ $N^*$ around 2040 MeV is needed to
fit the data~\cite{lism}.

The $p\bar n\pi^- + c.c.$ channel has also been studied from
$\psi^\prime$ decays~\cite{psip2}. The $N^*(1440)$ becomes the
largest signal and there are obvious structures for $M_{N\pi}>
2$~GeV in the $N\pi$ invariant mass spectra as shown in
Fig.~\ref{am}. But due to low statistics at BESII, no conclusive
information can be drawn for the $N^*$ resonances with mass above 2
GeV.

\begin{center}
\includegraphics[width=7.5cm]{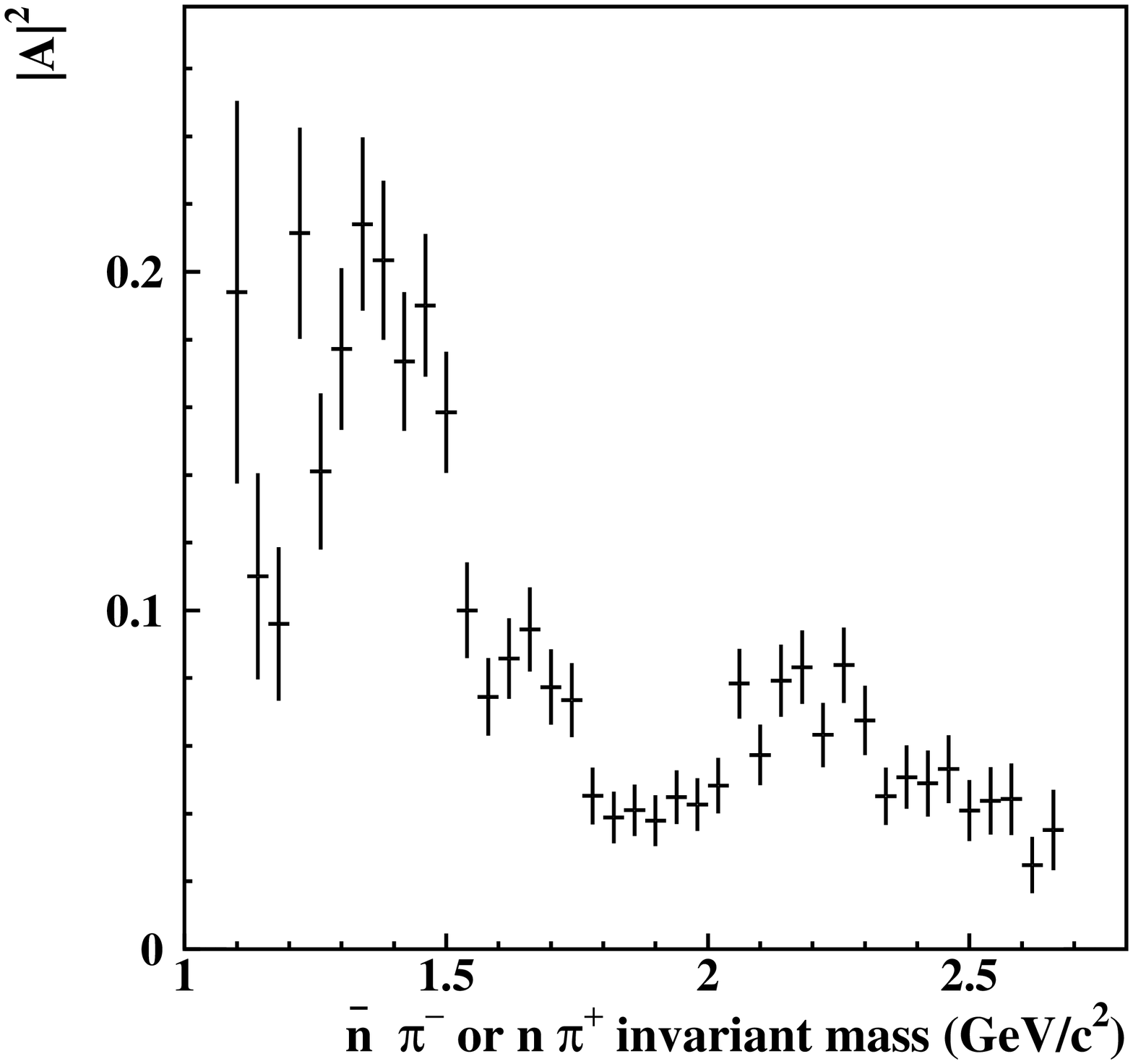}
\figcaption{\label{am}   Data divided by efficiency and phase space
vs  $\bar{n} \pi^-$ (or $n \pi^+$) invariant mass for $\psi' \to p
\bar{n} \pi^-+c.c.$ from Ref.[21].  }
\end{center}

Another very interesting result comes from the study of
$J/\psi\to\bar pp\eta$ and $J/\psi\to pK^-\bar\Lambda+c.c.$ channels
on the $N^*(1535)$ resonance. In $J/\psi\to\bar pp\eta$, as
expected, the $N^*(1535)$ gives the largest
contribution~\cite{ppeta}. In $J/\psi\to pK^-\bar\Lambda+c.c.$, a
strong near-threshold enhancement is observed for $K\Lambda$
invariant mass spectrum~\cite{Yanghx} as duplicated in
Fig.~\ref{fig:1}. The $K\Lambda$ threshold is 1609 MeV. The
near-threshold enhancement is confirmed by $J/\psi\to
nK_S\bar\Lambda+c.c.$~\cite{weidh}. Since the mass spectrum divided
by efficiency and phase space peaks at threshold, it is natural to
assume it comes from the sub-threshold nearby $N^*(1535)$ resonance.
Then from BES measured branching ratios of $J/\psi\to\bar
pp\eta$~\cite{ppeta} and $\psi\to
pK^-\bar\Lambda+c.c.$~\cite{Yanghx}, the ratio between effective
coupling constants of $N^*(1535)$ to $K\Lambda$ and $N\eta$ is
deduced to be around 1~\cite{lbc}. Recently, by treating the peak as
dynamically generated with unitary chiral theory, then the peak is a
coherent effect of $N^*(1535)$ pole and background,  and the ratio
between effective coupling constants of $N^*(1535)$ to $K\Lambda$
and $N\eta$ is deduced to be around 0.6~\cite{geng-oset}.

With previous known value of $g_{N^*(1535)N\eta}$, the obtained new
value of $g_{N^*(1535)K\Lambda}$ is shown to reproduce recent $pp\to
pK^+\Lambda$ near-threshold cross section data~\cite{cosy} as well.
There are also indications for the large $g_{N^*(1535)K\Lambda}$
from partial wave analysis of $\gamma p\to K\Lambda$
reactions~\cite{Lee}, the large $g_{N^*(1535)N\eta^\prime}$ coupling
from $\gamma p \to p\eta^\prime$ reaction at CLAS~\cite{etap} and
from $pp\to pp\eta^\prime$ reaction~\cite{caoxu1}, and large
$g_{N^*(1535)N\phi}$ coupling from $\pi^- p \to n\phi$, $pp\to
pp\phi$ and $pn\to d\phi$ reactions~\cite{xiejj,caoxu2}, but smaller
coupling of $g_{N^*(1535)K\Sigma}$ from comparison of $pp\to p
K^+\Lambda$ to $pp \to p K^+\Sigma^0$~\cite{Sibir}.

\begin{center}
\includegraphics[width=7.5cm]{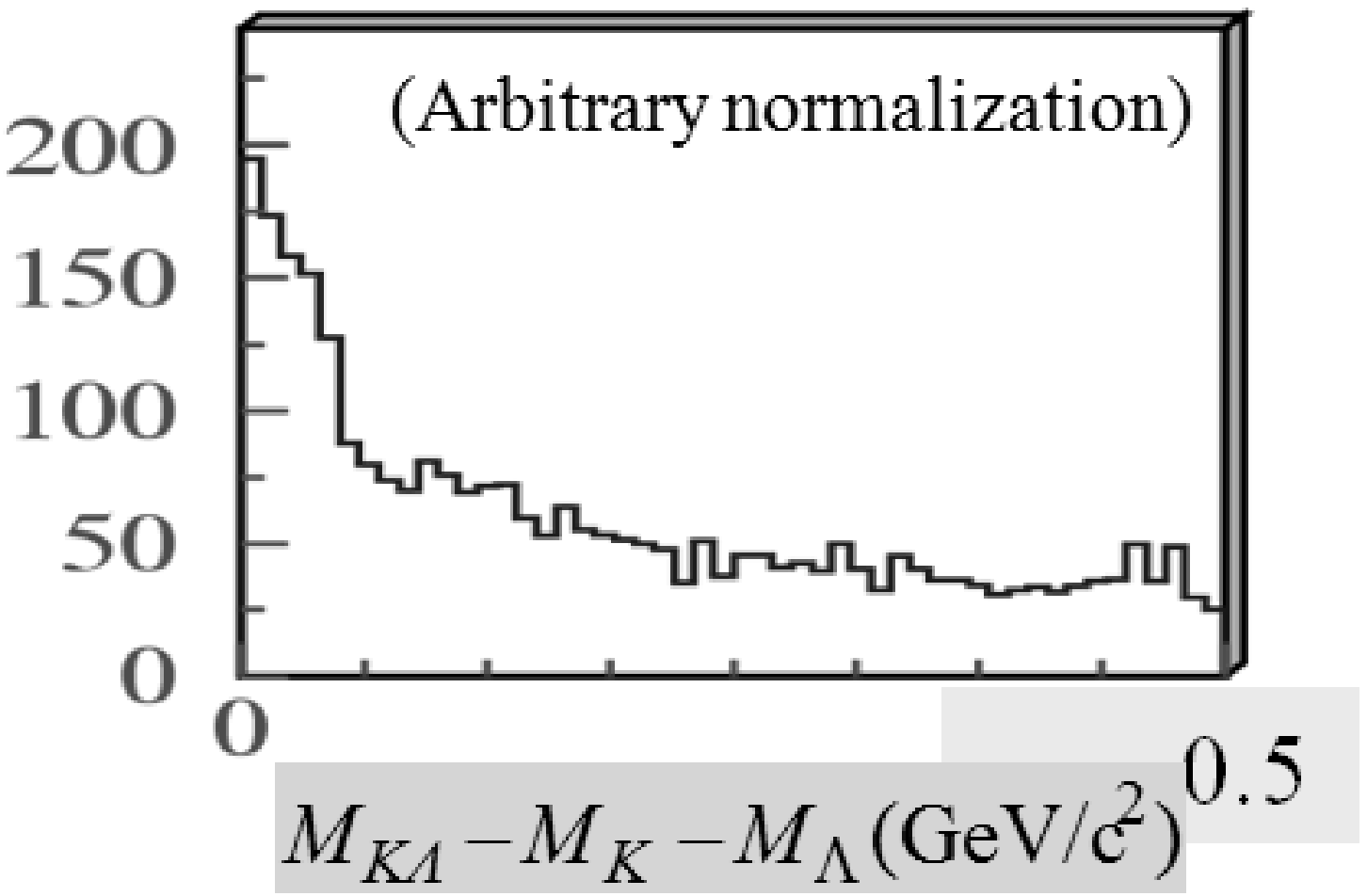}
\figcaption{\label{fig:1}  Invariant mass spectrum divided by
efficiency and phase space vs $M_{K\Lambda}\!-\!M_K\!-\!M_\Lambda$
for $J/\psi\to pK^-\bar\Lambda+c.c.$ from Ref.[23]. }
\end{center}

The observed decay pattern of the $N^*(1535)$ supports the picture
that there is a large mixture of the $|[ud][us]\bar s>$ pentaquark
component in the $N^*(1535)$. It not only gives a natural
explanation of the mass reverse problem of the lowest excited states
but also explains naturally its large couplings to the $N\eta$,
$N\eta^\prime$ and $K\Lambda$ meanwhile small couplings to the
$N\pi$ and $K\Sigma$. In the decay of the $|[ud][us]\bar s>$
pentaquark component, the $[ud]$ diquark with isospin $I=0$ is
stable and keeps unchanged while the $[us]$ diquark is broken to
combine with the $\bar s$ to form either $K^+(u\bar
s)\Lambda([ud]s)$ or $\eta(s\bar s)p([ud]u)$.

If this picture of large 5-quark mixture is correct, there should
also exist the SU(3) nonet partners of the $N^*(1535)$ and
$\Lambda^*(1405)$, {\sl i.e.}, an additional $\Lambda^*~1/2^-$
around 1570 MeV, a triplet $\Sigma^*~1/2^-$ around 1360 MeV and a
doublet $\Xi^*~1/2^-$ around 1520 MeV~\cite{zhu}. There is no hint
for these baryon resonances in the PDG tables \cite{PDG08}. However,
as pointed out in Ref.~\cite{Zou-Charm06}, there is in fact evidence
for all of them in the data of $J/\psi$ decays. According to
PDG~\cite{PDG08}, the branching ratios for
$J/\psi\to\bar\Sigma^-\Sigma^*(1385)^+$ and
$J/\psi\to\bar\Xi^+\Xi^*(1530)^-$ are $(3.1\pm 0.5)\times 10^{-4}$
and $(5.9\pm 1.5)\times 10^{-4}$, respectively. These two processes
are SU(3) breaking decays since $\Sigma$ and $\Xi$ belong to SU(3)
$1/2^+$ octet while $\Sigma^*(1385)$ and $\Xi^*(1530)$ belong to
SU(3) $3/2^+$ decuplet. Comparing with the similar SU(3) breaking
decay $J/\psi\to\bar p\Delta^+$ with branching ratio of less than
$1\times 10^{-4}$ and the SU(3) conserved decay $J/\psi\to\bar
pN^*(1535)^+$ with branching ratio of $(10\pm 3)\times 10^{-4}$, the
branching ratios for $J/\psi\to\bar\Sigma^-\Sigma^*(1385)^+$ and
$J/\psi\to\bar\Xi^+\Xi^*(1530)^-$ are puzzling too high. A possible
explanation for this puzzling phenomena is that there were
substantial components of $1/2^-$ under the $3/2^+$ peaks but the
two branching ratios were obtained by assuming pure $3/2^+$
contribution. In fact, a recent re-examination of some old data of
the $K^-p \to \Lambda\pi^+\pi^-$ reaction reveals that besides the
well established $\Sigma^{*}(1385)$ with $J^P=3/2^+$, there is
indeed some evidence for the possible existence of a new
$\Sigma^{*}$ resonance with $J^P=1/2^-$ around the same mass but
with broader decay width. This possibility could also be easily
checked with the high statistics BESIII data in near future.

With $10^9$ $\psi^\prime(3686)$ and $10^{10}$ $J/\psi$ events at
BESIII, the $N^*$, $\Delta^*$, $\Lambda^*$, $\Sigma^*$ and $\Xi^*$
can be well explored for masses up to 2740 MeV, 2450 MeV, 2570 MeV,
2490 MeV and 2360 MeV, respectively. Not only $J/\psi$ and
$\psi^\prime$ but also $\chi_{cJ}$ can have enough statistics for
studying these baryon resonances. Because the $\chi_{cJ}$ cannot
decay to hadrons through one virtual photon as vector charmonia do,
the $\chi_{cJ}$ decays provide an even better isospin filter for
studying baryon resonances.

\section{$N^*$ from $pp\to N N^*$}

The proton beams at COSY/Juelich and CSR/Lanzhou can provide $pp$
collisions with center-of-mass (CM) energies up to 3 GeV. The $pp\to
N N^*$ reaction can provide another useful source of information on
$N^*$ resonances. Many results from baryonic channels in charmonium
decays can be cross-checked by corresponding channels from $pp$
collisions. For example, comparing $J/\psi\to\bar p K^+\Lambda$ with
$pp\to p K^+\Lambda$, they share the same $K^+\Lambda$ resonances
and the same t-channel exchange interaction for $\bar p\Lambda$ and
$p\Lambda$. The large $N^*(1535)\Lambda K$ coupling observed in
$J/\psi\to\bar p K^+\Lambda$ should also have some reflection in
$pp\to p K^+\Lambda$, for which some very precise near-threshold
data are now available from COSY experiments~\cite{cosy,cosy2}.
Indeed a theoretical prediction without including the $N^*(1535)$
contribution~\cite{Tsushima} is obviously underestimating the
near-threshold data of COSY as shown by the dotted line in
Fig.~\ref{result}. After adding the contribution from the
$N^*(1535)$ with its coupling to $K\Lambda$ determined from $J/\psi$
decays~\cite{lbc}, the data can be reproduced perfectly as shown by
the solid line in Fig.~\ref{result}. While the $p\Lambda$ final
state interaction (FSI) is pointed out to play important role to
reproduce the cross section data~\cite{sib2}, the Dalitz plot
data~\cite{cosy} clearly show that both $p\Lambda$ FSI and
$N^*(1535)$ contribution are important.

\begin{center}
\includegraphics[width=7.5cm]{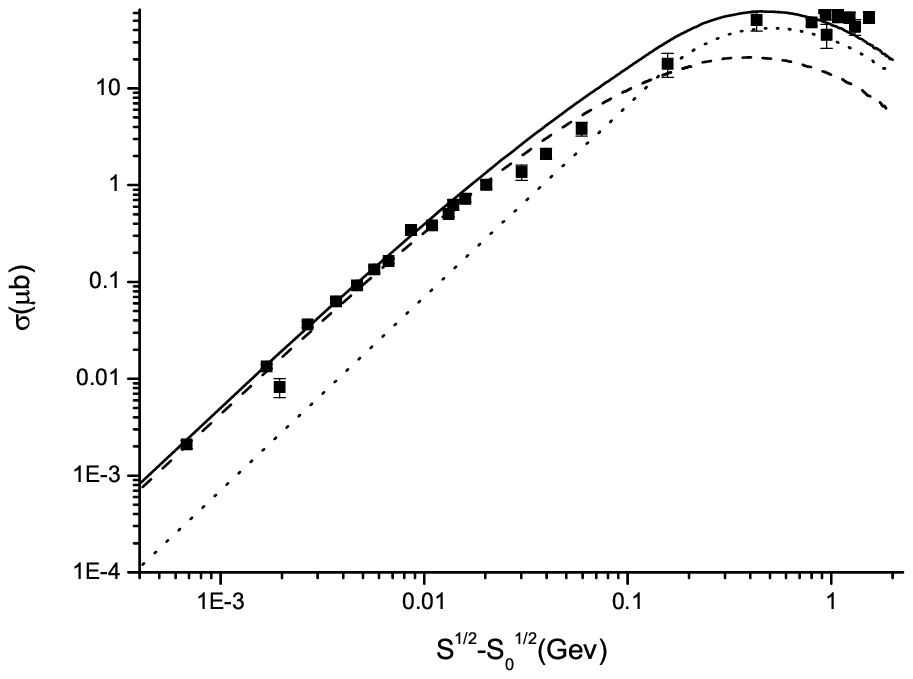}
\figcaption{\label{result}  The cross section of the reaction $pp\to
pK^+\Lambda$ as a function of the excess energy without (dotted
line) and with (solid line) including the contribution from
$N^*(1535)$ compared with data.  From Ref.[25]. }
\end{center}

The $N^*(1440)$ peak in the $n\pi^+$ invariant mass spectrum
observed in the $J/\psi\to\bar pn\pi^+$ reaction~\cite{bes2} is also
observed in the corresponding $pp\to pn\pi^+$ reaction by the
CELSIUS-WASA Collaboration~\cite{clement}. It is found that the
t-channel $\sigma$-meson exchange plays dominant role for the
production of the $N^*(1440)$ resonance~\cite{ouyang}. This suggests
that the $pp\to N N^*$ reaction is a good place for looking for
those ``missing" $N^*$ with large coupling to $N\sigma$. The $pp\to
pn\pi^+$ reaction at higher energies should be explored at COSY and
CSR.

Recently, the CELSIUS-WASA Collaboration observed an s-channel
resonance-like structure around 2.36 GeV in the $pn\to d\pi^0\pi^0$
reaction~\cite{WASA2}. It is just around $NN^*(1440)$ threshold.
Note that the $N^*(1440)$ has the same quantum number of nucleon and
has a large coupling to $N\sigma$. It is likely to form a
$NN^*(1440)$ quasibound state by t-channel $\sigma$ and other meson
exchanges as deuteron as a bound state of $pn$. Then the
$NN^*(1440)$ quasibound state decays in to $d\sigma$ due to the
large $N^*(1440)N\sigma$ coupling.

Besides the complementary study on the isospin 1/2 $N^*$ resonances,
the $pp$ collisions can also provide a new excellent source for
studying their isospin 3/2 partners, {\sl i.e.}, $\Delta^{++*}$
resonances. The spectrum of isospin 3/2 $\Delta^{++*}$ resonances is
of special interest since it is the most experimentally accessible
system composed of 3 identical valence quarks. However, our
knowledge on these resonances mainly comes from old $\pi N$
experiments and is still very poor~\cite{PDG08}. A recent
study~\cite{xiejj2} on $pp \to nK^+\Sigma^+$ reaction suggests that
the reaction is an excellent place for looking for those ``missing"
$\Delta^{++*}$ with large coupling to $p\rho^+$.

\begin{center}
\includegraphics[width=7.5cm]{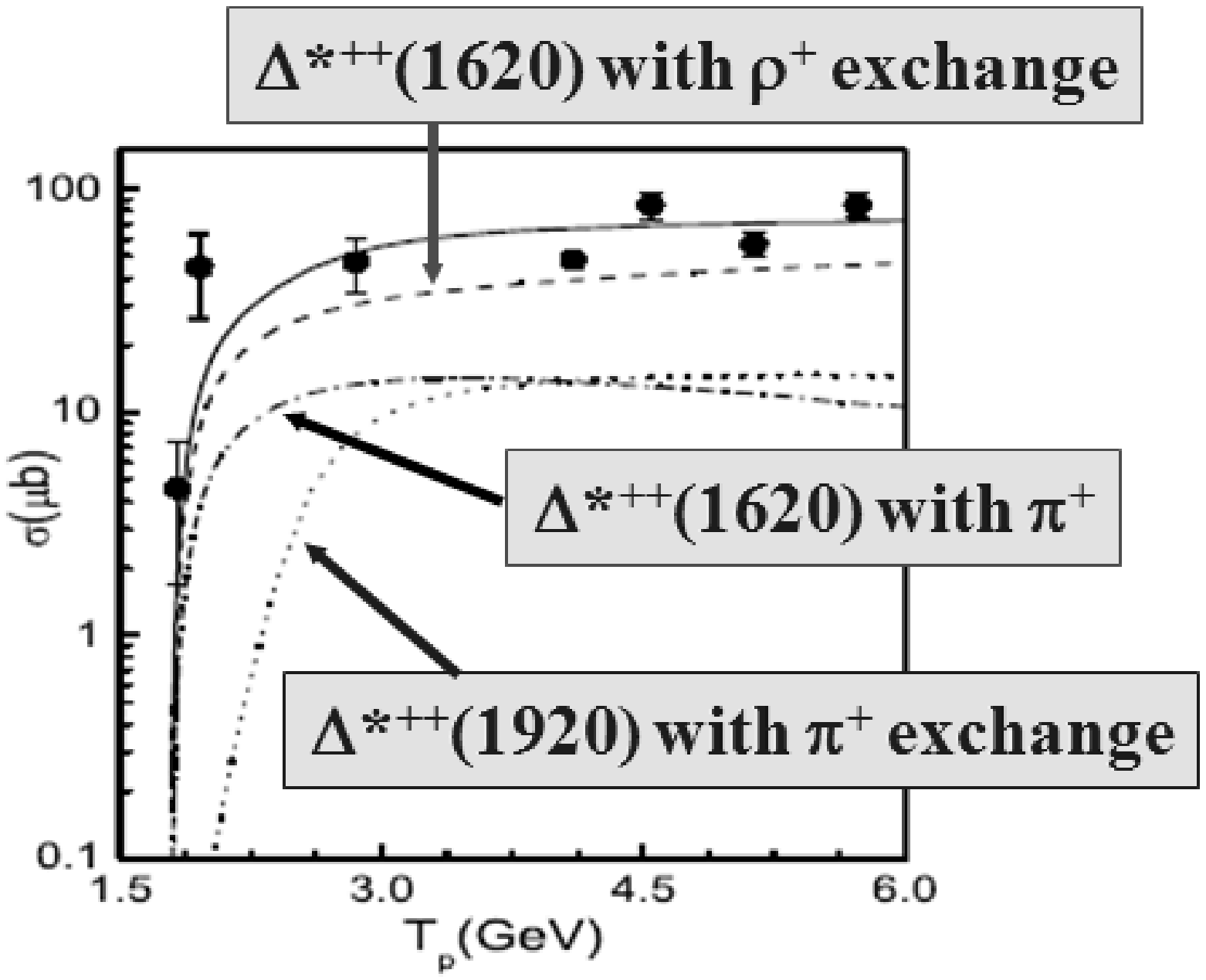}
\figcaption{\label{fig:2} Total cross section vs kinetic energy of
proton beam for the $pp \to nK^+\Sigma^+$ reaction: data [41,42] and
calculation (solid curve for sum of other curves) [40]. }
\end{center}

At present, little is known about the $pp\to nK^+\Sigma^+$ reaction.
Experimentally there are only a few data points about its total
cross section versus energy~\cite{data,06cosy11}. Theoretically a
resonance model with an effective intermediate $\Delta^{++*}(1920)$
resonance~\cite{tsushima} and the J\"{u}lich  meson exchange
model~\cite{gas} reproduce the old data at higher beam
energies~\cite{data} quite well, but their predictions for the cross
sections close to threshold fail by order of magnitude compared with
a recent COSY-11 measurement~\cite{06cosy11}. Recently this reaction
was restudied \cite{xiejj2}. With an effective Lagrangian approach,
contributions from a previous ignored sub-$K^+\Sigma^+$-threshold
resonance $\Delta^{++*}(1620)1/2^-$ are fully included in addition
to those already considered in previous calculations. It is found
that the $\Delta^{++*}(1620)$ resonance gives an overwhelmingly
dominant contribution for energies very close to threshold, with a
very important contribution from the t-channel $\rho$ exchange as
shown in Fig.~\ref{fig:2}. This may solve the problem that all
previous calculations seriously underestimate the near-threshold
cross section by order(s) of magnitude.

A important implication of this study is that the $pp\to
n\Delta^{++*}$ may provide a good source for exploring
$\rho^+p\to\Delta^{++*}$ and should be further studied at COSY and
CSR.

A more recent measurement of the $pp\to nK^+\Sigma^+$ reaction near
its threshold by ANKE collaboration~\cite{ANKE} gives a much smaller
cross section than those by COSY-11. This would mean much smaller
$\Delta^{++*}(1620)$ contribution and $n\Sigma^+$ FSI. Since both
detectors are not $4\pi$ solid angle detectors, there is model
dependence to deduce the total cross section from a fraction of
$4\pi$ solid angle measurement. A good Dalitz plot measurement with
a good $4\pi$ solid angle detector would be very helpful to settle
down the contradiction.

\section{$N^*$ from $\bar pp\to\bar N N^*$}

The antiproton beam at PANDA/FAIR is going to perform $\bar pp$
collision experiment with beam momenta ranging from 1.5 to 15 GeV.
The $\bar pp$ collisions could provide a much richer source for the
production of baryon resonances than $e^+e^-$ collisions. All the
final states of $e^+e^-$ collisions and much more other states are
accessible by $\bar pp$ collisions. A large portion of $\bar pp$
final states contain baryons and should not be wasted at PANDA/FAIR.

Recently, a proposal is made to study $N^*$ resonances with $\bar
pp\to\bar pn\pi^+$ reaction~\cite{Wujiajun}. Due to absence of the
$\Delta^{++}$ production for this reaction, the contribution of the
$\Delta$ excitation is much smaller than in the corresponding $pp
\to pn\pi^+$ reaction. It is found that for the beam momenta around
$1.5\sim 3$ GeV, the contribution of the Roper resonance $N^*(1440)$
produced by the t-channel $\sigma$ exchange dominates over other
contributions due to its known large coupling to
$N\sigma$~\cite{PDG08,hir}, as shown by the predicted $n\pi^+$
invariant mass spectrum for the reaction at $T_{\bar p}=2.88$ GeV in
Fig.\ref{fig:4}. This will provide the cleanest place for studying
the properties of the Roper resonance and the best place for looking
for other ``missing" $N^*$ resonances with large coupling to
$N\sigma$.

\begin{center}
\includegraphics[width=7.5cm]{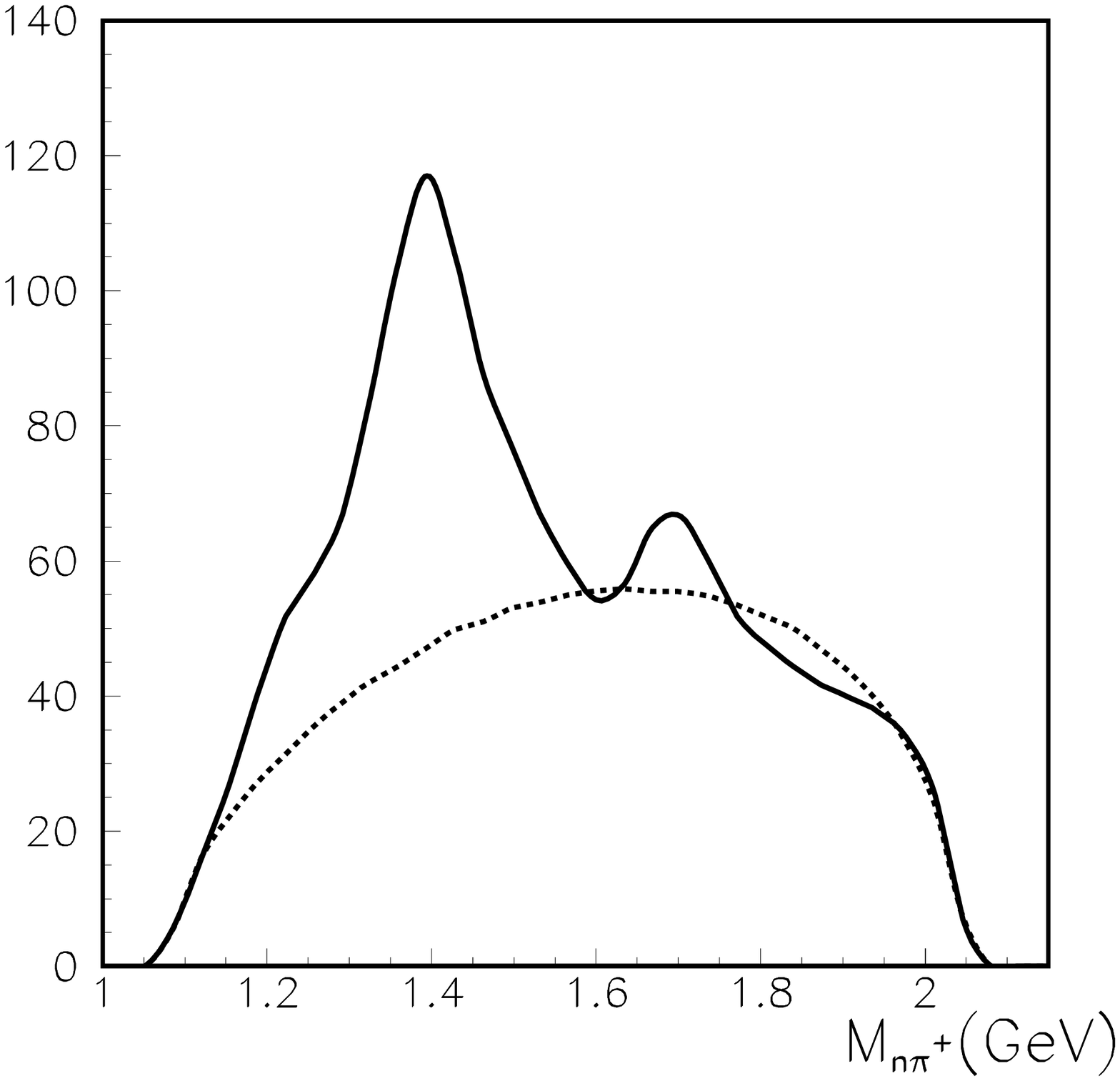}
\figcaption{\label{fig:4} Predicted $n\pi^+$ invariant mass spectrum
(solid curve) for the $\bar{p}p \to \bar{p} n \pi^+$ reaction at
$T_{\bar p}=2.88$ GeV, compared with phase space distribution
(dashed curve) [46]. }
\end{center}

Another interesting possibility is that the poorly known $\Omega^*$
resonances may be produced and studied by PANDA/FAIR experiment
while they cannot be studied from charmonium decays due to limited
energy.

\section{Summary and prospects}

In summary, complementary to the conventional experimental studies
on $N^*$ from $\pi N$ and $\gamma^{(*)}N$ reactions, the $e^+e^-$,
$pp$ and $p\bar p$ collisions can give novel insights into these
$N^*$ resonances.

The $e^+e^-$ collisions through production and decay of charmonia
$\psi$ and $\chi_{cJ}$ provide a nice isospin filter for a
simultaneously study of $N^*$, $\Delta^*$, $\Lambda^*$, $\Sigma^*$
and $\Xi^*$. With $10^9$ $\psi^\prime(3686)$ and $10^{10}$ $J/\psi$
events at BESIII, the $N^*$, $\Delta^*$, $\Lambda^*$, $\Sigma^*$ and
$\Xi^*$ can be well explored for masses up to 2740 MeV, 2450 MeV,
2570 MeV, 2490 MeV and 2360 MeV, respectively. Many new baryon
resonances should be observed.

The $pp$ collisions should be the best place for producing those
$\Delta^{*++}$ with large coupling to $\rho^+p$ though $pp\to
n\Delta^{*++}$ reaction. It is also a nice place for studying the
$N^*$ resonances with large coupling to $N\sigma$. The COSY/J\"ulich
is short of a good $4\pi$ solid angle detector for both charged and
neutral particles for a comprehensive study of the baryon spectrum.
The study of this aspect should be continued at Lanzhou CSR with the
schedule $4\pi$ solid angle detector HPLUS for both charged and
neutral particles~\cite{xuhs}.

The $p\bar p$ collisions should be the best place for looking for
those $N^*$ with large coupling to $\sigma N$ and for the study of
the poorly known $\Omega^*$ resonances. A large portion of $\bar pp$
final states contain baryons and should not be wasted at the
forthcoming PANDA/FAIR experiment with the antiproton beam. Instead
PANDA should play important role on baryon spectroscopy.

With $e^+e^-$ experiment at BESIII/BEPCII, $pp$ experiment at
HPLUS/Lanzhou, $\bar pp$ experiment at PANDA/FAIR joining the force
of $\gamma^{(*)}$ experiments at CEBAF/JLAB, ELSA, Spring-8, and $K$
beam experiment at JPARC for the study of baryon spectrum, a new era
of baryon spectrum study is foreseeing to come.

\vskip 1cm

\acknowledgments{I would like to thank Bo-chao Liu, Ju-jun Xie,
Jia-jun Wu, Huan-ching Chiang, Peng-nian Shen, Jian-xiong Wang, Zhen
Ouyang, Xu Cao, Hu-shan Xu and my BES collaborators for
collaboration on relevant issues presented here.}

\end{multicols}

\vspace{-2mm}
\centerline{\rule{80mm}{0.1pt}}
\vspace{2mm}

\begin{multicols}{2}

\end{multicols}

\clearpage

\end{document}